\documentclass[11pt]{article}
\usepackage{cospar}
\usepackage{url}
\usepackage{graphics}
\usepackage[figuresright]{rotating}

\usepackage{graphicx}
\usepackage[figuresright]{rotating}

\title{SOLAR FLARE ELECTRON ACCELERATION: COMPARING THEORIES AND OBSERVATIONS}

\author{A. O. Benz and P. Saint-Hilaire \vskip-1.5cm \address{Institute of Astronomy, ETH, 8092 Zurich, Switzerland}}

\begin{document}

% typeset front matter
\maketitle

\begin{abstract}
A popular scenario for electron acceleration in solar flares is transit-time damping of low-frequency MHD waves excited by reconnection and its outflows. The scenario requires several processes in sequence to yield energetic electrons of the observed large number. Until now there was very little evidence for this scenario, as it is even not clear where the flare energy is released. RHESSI measurements of bremsstrahlung by non-thermal flare electrons yield energy estimates as well as the position where the energy is deposited. Thus quantitative measurements can be put into the frame of the global magnetic field configuration as seen in coronal EUV line observations. We present RHESSI observations combined with TRACE data that suggest primary energy inputs mostly into electron acceleration and to a minor fraction into coronal heating and primary motion. The more sensitive and lower energy X-ray observations by RHESSI have found also small events (C class) at the time of the acceleration of electron beams exciting meter wave Type III bursts. However, not all RHESSI flares involve Type III radio emissions. The association of other decimeter radio emissions, such as narrowband spikes and pulsations, with X-rays is summarized in view of electron acceleration.
\end{abstract}

\section*{INTRODUCTION}
\vskip2mm
There are many theories and proposals for the particle acceleration processes involved in flares. Over the past twenty years, however, most have focused on the process of magnetic reconnection. The most popular classes of mechanisms for electrons are acceleration by an electric field parallel to the magnetic field, by shocks and stochastic acceleration by waves (Miller et al., 1997, and references therein). Considering the large number of non-thermal electrons required by thick target emission, the first two processes become questionable for interpreting the observed hard X-ray flux at 15--500 keV. A popular version of stochastic acceleration (second-order Fermi acceleration) is transit-time damping, in which thermal electrons are in quasi-resonance with magnetoacoustic waves (Fisk, 1976; Stix, 1992). After or during these primary processes, accelerated electrons may become unstable to wave generation, escape and may produce beam emissions, such as thin target bremsstrahlung and Type III radio bursts (bump-on-tail instability driving Langmuir waves). They may get trapped in coronal magnetic fields and finally precipitate to produce thick target emission.

\begin{figure}
%\vskip-3.5cm
\centerline{
\includegraphics[width=75mm]{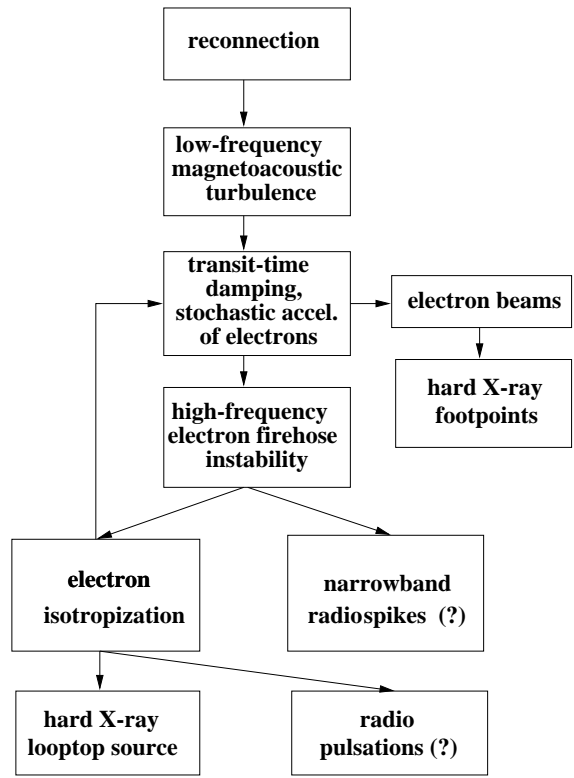}}
\caption[fig1]{Schematic sequence of transit-time electron acceleration in a solar flare. Possible observable radiations are indicated, including radio emissions tentatively associated with acceleration.}
\end{figure}

For the first time there is a complete scenario that can explain flare electron acceleration by a self-consistent chain of sub-processes (Fig.1). It includes the following four steps: {\sl (i)} Reconnection triggers the simplification of magnetic field lines to an energetically lower geometry. {\sl (ii)} In the reconnection jets where most of the free energy is initially released, MHD turbulence develops including magnetoacoustic waves. {\sl (iii)} These wave are damped by thermal electrons that are accelerated to mostly subrelativistic speed. {\sl (iv)} The accelerated electron velocity distribution has an enhanced tail in parallel direction and possibly other deviations from a simple maxwellian. Transit-time acceleration has recently been complemented with an investigation on the instability of the accelerated electrons by Paesold \& Benz (1999), who demonstrated the Electron Firehose Instability to set in. Its main effect, however, is not energy loss but scattering the electrons into more transverse velocities and hereby allowing further  acceleration (Paesold \& Benz, 2003). Thus the electrons are expected to drive velocity space instabilities in the course of acceleration. {\sl (v)} Finally, the electrons escape from the acceleration region and emit secondary emissions by forming a beam, getting trapped, or hitting the chromosphere as a thick target. A schematic representation of this scenario is given in Fig. 1. 

Here we ask the question: Which observations support this scenario? In particular, we concentrate on the new observations by the Reuven Ramaty High Energy Solar Spectroscopic Imager (RHESSI) and use them to test the sub-processes presented above.

\section*{RECONNECTION AND ENERGY PARTITION}
\vskip2mm
Magnetic reconnection in solar flares is supported by observations of soft X-ray flare emission forming a cusp shape structure. More recently, inflow perpendicular to the magnetic field into the reconnection region has been reported by Yokoyama et al. (2001). The identification of reconnection in the corona was often based on the presence of dense material that became visible in the course of a flare. As the process of reconnection is not predicted to substantially increase the density, the region seems to have been brightened by evaporated material, which then outlined a geometry suggestive of magnetic reconnection. Thus cusp-shaped features have become questionable indicators for reconnection (Grechnev \& Nakajima, 2002).

\begin{figure}
\centerline{
\includegraphics[width=95mm]{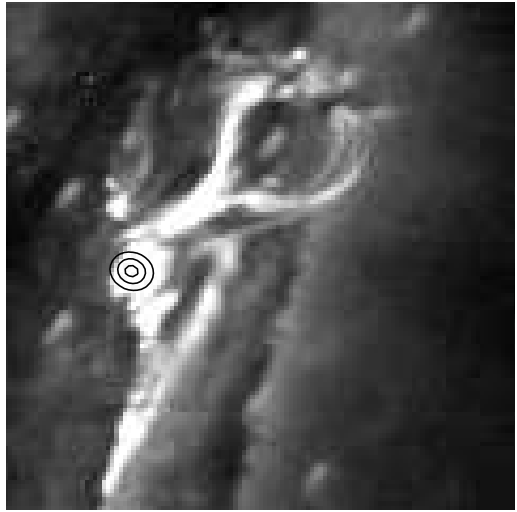}}
%\vskip9cm
\caption[fig2]{Solar flare on 26 February 2002, 10:28 UT. A TRACE image at 195\AA\ is overlaid with RHESSI observations (isophotes) of 12--25 keV emission during the impulsive phase (from Saint-Hilaire \& Benz, 2002).}
\end{figure}

Observations indicative of reconnection jets have been noticed by Shibata et al. (1994), Pohjolainen et al. (2001), Zang, Wang \& Liu (2000) and others. Bulk motions of relatively cool material are often seen before the start of the HXR event. Innes et al. (1997) have reported reconnection jets in the quiet Sun having a temperature of a few $10^5$K. MHD theory of reconnection predicts equal shares of energy for local heating by electric resistivity and the motion of the plasma ejected from the reconnection region (e.g. Priest \& Forbes, 2002). Thus the bulk kinetic energy involved in the jets is a primary form of released energy and is predicted to be an important energy input into the corona.

The initial partitioning of the released energy is not well known, as in reality a considerable fraction of the flare energy is initially transferred into energetic electrons (Neupert, 1968; Brown, 1971; Lin \& Hudson, 1976). These 'non-thermal' electrons carry a large fraction of the energy away from the flare region and deposit it partially in the chromosphere, where plasma is heated to tens of million degrees. The plasma rises into the corona and emits thermal soft X-rays (SXR), as first described by Neupert (1968). 
	
With the new generation of solar instruments in EUV lines, soft and hard X-rays, a more accurate determination of the various forms of energy becomes feasible. Recently, Saint-Hilaire \& Benz (2002) have presented an event observed by TRACE at 195 \AA\  and RHESSI (Fig.2). The TRACE observations, showing a blend of relatively cool material in Fe XXII and possibly hot plasma in Fe XXIV, display an ejection that does not leave the Sun, but apparently transfers its kinetic bulk energy into the corona. The ejection is interpreted as a reconnection jet, as it starts simultaneously with the HXR emission. The motion observed is outward. No inward motion is observed that would indicate the other reconnection jet. However, the region below the suspected X-point is small and very bright in EUV so that it would be difficult to notice. RHESSI observed a HXR source close to the place where the downward jet would have been expected. This is also the site of a cusp-shaped structure visible brightly in EUV. Also in this region appears a thermal SXR source as observed by RHESSI below 15 keV. Its temperature is 20.8($\pm$0.9) MK (isothermal fit).

\begin{figure}
\centerline{
\includegraphics[width=100mm]{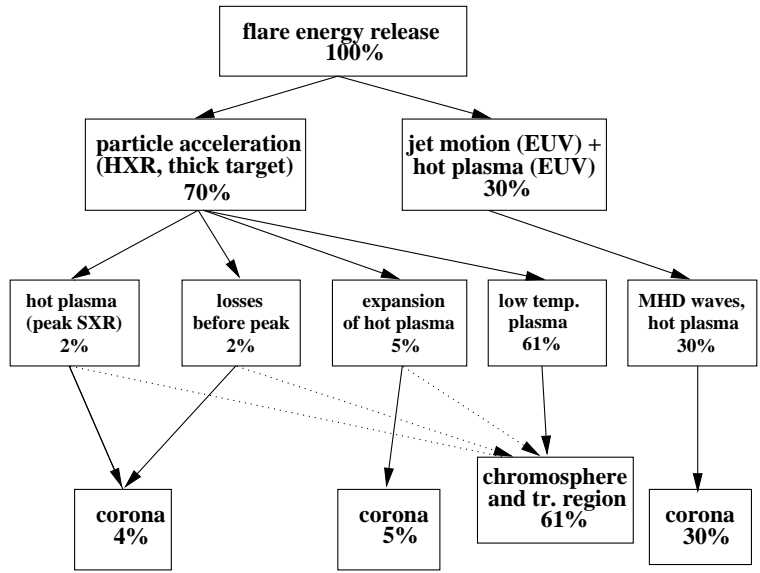}}
%\vskip10.0cm
\caption[fig3]{Schematic distribution and flow of energy in the flare of 26 February 2002. Approximate (!) energy partition in percentage of the total is indicated as observed in the flare well observed by RHESSI and TRACE (from Saint-Hilaire \& Benz, 2002).}
\end{figure}

The above interpretations suggest that the flare energy initially went into electron acceleration, plasma motion (reconnection jet), and heating of the ejected material. Other forms, such as ion acceleration, are not observable in this flare and are neglected. The kinetic energy of the precipitating electrons is assumed to thermalize in the chromosphere and is estimated from a thick-target model. Their bremsstrahlung radiation is observed in HXR emission exceeding 10 keV. Plasma motion is observed and measured in EUV (TRACE observations). The thermal energy of the ejecta is also estimated from EUV observations. The observed total energy release amounts to 3.7$\times 10^{30}$erg. The second line in Fig. 3 displays the primary energy partition and includes the best-estimate percentages as measured by Saint-Hilaire \& Benz (2002). The energy in the possible reconnection jet motion is only 3\% of the total, thus apparently an order of magnitude less than in the hot plasma. Note that the partitions into the various forms of energy are tentative, particularly for the part that stays in the corona, namely the energy of the bulk motion of a possible reconnection jet and the thermal energy content of the heated material that was in the corona before and after the flare. They must be considered order-of-magnitude estimates, rounded such that each line adds up to 100\%. 

The secondary forms of energy are listed in the third line of Fig. 3. The thermal energy content of the flare kernel was measured in the 10--15 keV range by a fit to the RHESSI spectrum at low-energy peak time (Fig.2). The thermal energy content can be compared to the thermal radiation emitted by this plasma over the entire event, calculated from the measured flux in the observed band and extrapolated to the entire wavelength range using the observed temperature and emission measure. The radiated energy amounts to only 5\% of the thermal energy content (Saint-Hilaire \& Benz, 2002). It suggests that most of the thermal energy is lost in other ways, such as by heat conduction and emissions at lower temperature.

The third line of Fig. 3 also includes other forms of energy inferred from the thermal energy content measured at peak soft X-rays (SXR, $<$ 15 keV). The other forms have been estimated for nanoflares in the quiet corona by Benz \& Krucker (2002), who suggested that {\sl (i)} the total energy content must be doubled, as roughly the same energy is lost before that peak as afterwards, and that {\sl (ii)} the expansion of the heated plasma into the corona absorbs about three times as much energy. The rest of the energy deposited by the electron beam heats low temperature plasma in the transition region and chromosphere. For the flares considered here, the energy loss before the peak is less as regular SXR flares are usually asymmetric having a longer decay time; on the other hand flare loops are larger and require more energy for the expansion. As the different partitions between nanoflares and regular flares are within the accuracy of these estimates, we have used here the nanoflare values.

Keeping in mind the large uncertainties given in Fig. 3, we note nevertheless that relatively little of the initial energy in accelerated electrons is observed in soft X-ray emission. Even if this value is corrected for energy that is lost by radiation and conduction before being measured at SXR peak time, and by the energy needed for expansion into the corona, less than 10\% of the total flare energy seem to end up in the hot flare plasma of the kernel. The rest, more than 60\%, gets apparently lost in the chromosphere and transition region. Thus the conversion of non-thermal energy into hot thermal plasma seems to be less than 20\% efficient in the observed flare.

The bottom line of Fig. 3 displays where the energy ends up. Roughly 40\% of the released flare energy appears in thermal form of a plasma at coronal temperature.  Some of the heated plasma ultimately cools to temperatures below the observing range, where its remaining energy is lost from the point of view of coronal observations. Contrary to ananflares, this remaining energy is realtively small. We estimate it from the ratio of the minimum coronal temperature to the flare temperature, thus about 0.05.

\section*{STOCHASTIC ACCELERATION}
\vskip2mm
Stochastic acceleration can be modeled by a diffusion of the particles in energy $E$,

\begin{equation}
\left( {\partial \over \partial t} + v{ \partial \over \partial z}\right) f(E,z) = {\partial \over \partial E}\left( D E \left( {\partial \over \partial t} {f(E,z) \over E^{1/2}} \right)\right)\ +\ \left( {\partial f \over \partial t}\right)_{\rm coll} \ \ \ .
\end{equation}

A spatial derivative along the magnetic field, assumed to be the $z$ direction, is included to account for particle loss. Let us assume for simplicity that the derivative of the electron distribution in energy, $f(E)$, can be approximated by a scale length $L$ and that the diffusion coefficient, $D$, is proportional to the wave energy density $W$. In steady state, Eq.(1) takes the form of a Bessel equation, having the solution

\begin{equation}
 f(E)\ =\ C_1 E^{-d+1/2}K_d(E)\ \ \ ,
\end{equation}
where $C_1$ is a constant and $K$ is the Macdonald function of order $d$, a combination of Bessel functions (Benz, 1977). The order $d$ again is a product of constants, and for $d\ll d_c$, a critical value, the solution has approximately the form of a power-law, 

\begin{equation}
\ f(E)\ \approx \ f_o E^{-\delta}\ \ \ ,
\end{equation}
where
\begin{equation}
f_o\ \sim\ \left(WL\right)^{7/8}
\end{equation}
and
\begin{equation}
\delta\ \sim\ \left(WL\right)^{-1/2} \ \ \ .
\end{equation}
Eqs. (4) and (5) predict that the number of accelerated electrons and their power-law index anti-correlate. Nevertheless, the above diffusion relation and its solution are general and apply to any type of waves, including magnetoacoustic waves in transit-time acceleration. 

The electron energy distribution has been converted into thick target bremsstrahlung flux, $I_{\rm HXR}$ by Brown \& Loran (1985). Eliminating $WL$ in Eqs.(4) and (5), they find 

\begin{equation}
I_{\rm HXR} \  = \ C_2 {{\left(1/2 + 1/2[1+4\alpha(\gamma+3/2)]^{1/2}\right)^2}\over{[(\gamma - 1)(\gamma + 3/2)]^2}}\ \ \ .
\end{equation}
where $C_2$ is a constant, $\gamma$ the power-law index of the observed X-ray spectrum, and $\alpha$ is the ratio of the acceleration region size to the mean free path.

Anti-correlation between the power-law spectral index of HXR photons and flux has been noticed since the early times of HXR flare observations. The 'soft-hard-soft' behavior has been reported Parks \& Winkler (1969) and is the most abundant type of flares (see also Dennis, 1985). Moreover, anti-correlation of power-law index and individual flare peaks has been noted in HXR and microwaves (Benz, 1977). RHESSI can observe this property with high time resolution as shown in Fig. 4. The correlation remains excellent when a minimum envelope is subtracted and only the fine time structures are considered.

Fletcher \& Hudson (2002) have related the anti-correlation to the flare geometry, invoking acceleration by a field aligned electric field. Thus the observed anti-correlation is not a unique feature of a particular acceleration mechanism. Nevertheless, we conclude that it is at least consistent with stochastic acceleration.

\begin{figure}
\centerline{
\includegraphics[width=120mm]{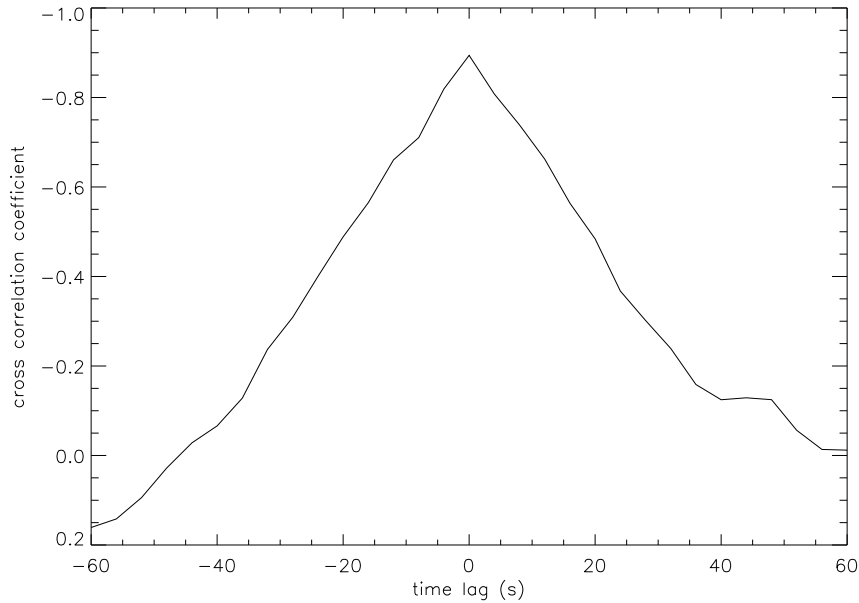}}
%\vskip7.7cm
\caption[fig4]{Cross-correlation between hard X-ray flux (30--50 keV) and spectral power-law index fitted in the same energy range in every time bin of 4 seconds. The enhanced HXR emission was observed by RHESSI on 2002 August 22, 01:53--57 UT (from Grigis \& Benz, 2003).}
\end{figure}

\section*{VELOCITY SPACE INSTABILITIES AND THE ABSENCE OF COHERENT RADIO EMISSION}
\vskip2mm
Stochastic acceleration of electrons requires waves. Some of the proposed waves, such as Langmuir waves proposed e.g. by Benz (1977) and Kuijpers et al. (1981), would produce strong radio emissions. They were later found to be difficult to excite (e.g. Vlahos \& Papadopoulos, 1982) and cannot be reconciled with the absence of strong radio emission in flares (see below). Another possible cause for radio emission are accelerated electrons, reaching a velocity distribution that is unstable to some wave mode. Many such situations, causing intense radio emission, are conceivable. Radio emissions from the acceleration regions of flares may thus be expected at characteristic frequencies, such as the plasma frequency or the electron gyrofrequency and their harmonics, all most likely at decimeter wavelengths. 

During flares the total solar radio emission in meter and decimeter waves occasionally brightens up by several orders of magnitude. These radio sources thus outshine the thermal radiation of the rest of the sun and are called bursts. The common characteristic of these emissions is that they are not produced by single electrons as in free-free emission (bremsstrahlung) or synchrotron radiation, but by waves and instabilities in the plasma. The wave introduces an organization in the plasma that allows particles to emit in phase, thus coherently. Coherent emissions are generally very efficient in converting kinetic particle energy into radiation. Thus coherent emissions are bright and indicate the presence of waves in the plasma. The waves that can give rise to escaping radio emissions must be of high frequency, above the local plasma frequency, and are generally excited by some instability. Thus coherent emissions indicate the presence of a cause for instability such as a beam of electrons, magnetically trapped electrons having a loss-cone velocity distribution, a shock, or an unstable current.

First spectral observations were made already in the early 1960s by Young et al.(1961), exploring the spectral region beyond 300 MHz up to 1 GHz. In addition to high-frequency Type III bursts, these observers noted a rich variety of 'Type III-like' features later termed pulsations and fine structures in Type IV bursts, such as drifting parallel bands or intermediate drift bursts. The decimeter range has been systematically studied only in the 1980s when digitally recording spectrometers became available resolving the fast temporal changes and narrowband spectral structure (G\"udel \& Benz, 1988). In the 1990s the spectral coverage was extended finally beyond 10 GHz where coherent emissions were found to become weak and rare (Isliker \& Benz, 1994; Bruggmann et al. 1990). Today several spectrometers observe in the decimeter range, and the spectral properties of coherent radiations have been well explored in the past decade. These observations have revealed a new field of solar emissions with potential diagnostic value for flare physics and particle acceleration.

Flare related radio emission is indeed observed, but not in every flare, as pointed out already by Simnett \& Benz (1986). The existence of flares without radio emissions excludes vigorous electron acceleration and suggests a more gentle extraction of electrons from the thermal population, such as proposed for instance by transit-time acceleration (Fig.1). As noted before, the electron anisotropy caused by preferential acceleration along the magnetic field may be susceptible to the Electron Firehose Instability (Paesold \& Benz, 1999). The non-resonant instability drives ion cyclotron waves. These waves have frequencies around the ion gyrofrequency and do not excite radio emission.

However, if electrons are accelerated to high energies, they become resonant with ion cyclotron waves and the Resonant Electron Firehose Instability may occur. Messmer (2002) showed that this instability causes the electrons to diffuse into a horse-shoe distribution which under certain conditions becomes unstable to Langmuir waves. Such waves are known to produce radio emission e.g. by wave-wave coupling (harmonic emission). They may be the cause of narrowband spikes occasionally observed in decimeter radiation of HXR flares. Alternatively, broadband drifting pulsations are sometimes observed in association with HXR flares (Kliem et al. 2000; Kahn et al. 2002).

There are several possible causes for the absence of radio emission: The radiation from the acceleration process may be beamed in a different direction or hidden by absorption, accelerated electrons may not develop an instability, and shocks may not become visible. At present it is not clear what coherent radio emissions originate at the time of HXR emission as a result of the acceleration process. The fact that there is sometimes no decimeter radiation at all, however, is consistent with the 'standard model' presented in Fig. 1 that can accelerate and isotropize electrons without leaving an observable decimeter radio signature.

\section*{ESCAPING BEAMS AND TRAPPING}
\vskip2mm
As soon as electrons are accelerated, they can produce hard X-ray emission, observable possibly as a loop-top source. Once electrons escape from the acceleration region, a beam distribution builds up in velocity space, unstable to the bump-on-tail distribution of Langmuir waves. In addition, electrons may be reflected on converging magnetic fields building up a loss-cone distribution unstable to upper hybrid waves. Both Langmuir waves and upper hybrid waves cause radio emission. 

In a recent study on 11 C-class flares in active region 9830, none showed a loop-top source in RHESSI observations. In 7 events, the Phoenix-2 radio spectrometer recorded radio emission between 100 MHz and 4 GHz, 5 of them Type III radio bursts. Two were associated with decimeter wave emissions of unknown origin (Benz \& Grigis, unpublished). Most surprising in our recent investigations is the large range of observed ratios between coherent radio emission and X-ray flux. The range exceeds 9 orders of magnitude from intense groups of meter wave Type III bursts without X-ray counterpart to large HXR events without coherent radio emission at meter and decimeter waves.

Again the absence of radio emission may have several causes. In addition to beaming and absorption, escaping electrons may not form an unstable beam. If the accelerator initially produces a power-law velocity distribution, electrons would need to travel a distance of several times $v_{te}\tau$ to become unstable, where $v_{te}$ is the mean thermal electron velocity and $\tau$ the e-folding acceleration time (Benz, 2002). Similarly, precipitating electrons may not form an unstable loss-cone because trapping can be inefficient if the density on the path is high and the collision rate too large. Alternatively, the electrons could be confined by closed magnetic fields or wave turbulence, such that they cannot propagate out of the acceleration region.

Both radio emission inhibiting reasons given above for electrons escaping the acceleration region apply best to acceleration sites at low altitude. Such a location is supported by observations like Fig. 2, in which no appreciable HXR emission was observed at higher altitudes where most of the plasma motion was observed.

\section*{CONCLUSIONS}
\vskip2mm
Observations at several wavelengths are found generally consistent with a standard flare model that avoids excessive radio emission. The radio emission can be completely absent in some cases. In others cases, such as often observed in the flares associated with interplanetary electron events (SEP), meter wave Type III radio bursts are the dominant emission. 

The proposed sequence of processes in a widely considered model, here termed 'standard', must and can be investigated individually. We find that none of the currently available observations contradict that model. This does not mean that it is proven for all cases. Particularly relevant observations would be the measurement of line emissions and Doppler shifts at very high temperatures expected in the turbulence of reconnection outflows, and the localization of coherent radio emissions.

RHESSI observations yield energy estimates and approximate energy partitions, indicating that the main flare energy first appears in the form of precipitating fast electrons. In a well observed flare, they seem to be released at relatively low altitude. TRACE images show motions suggestive of a reconnection jet, but only in the upward direction. The energy in this bulk motion was estimated to be smaller than the energy in the non-thermal electrons. The observations of one flare need to be corroborated by the study of a larger sample.

\vskip1cm
\noindent E-mail address of A. O. Benz: benz at astro.phys.ethz.ch\par
\noindent Manuscript received 29 November 2002; accepted 20 January 2003.

\end{document}